\documentclass[prd,showpacs,preprint,nofootinbib]{revtex4}
\usepackage{}
\usepackage{graphicx,color,amsmath}
\usepackage{bm}
\usepackage{slashed}
\usepackage{extarrows}
\allowdisplaybreaks
\newcommand{\D}{{\rm{d}}}

\usepackage{color}

\begin{document}
\title{One loop corrections to the trilinear self
coupling of the Higgs boson in the standard model with inclusion of singlet vector-like top quark}
\author{Jin Zhang\footnote{jinzhang@yxnu.edu.cn}}
\affiliation{Department of Physics, Yuxi Normal University,
Yuxi, Yunnan, 653100, China}


\begin{abstract}
The complete one loop corrections to the trilinear
self coupling of the Higgs boson is firstly evaluated
in the Standard Model (SM) by the effective potential approach.
Then we extend the SM by including a singlet vector-like
top partner and assume that the top partner only
mixes with the top quark, the effects of the mass of the top partner and the
mixing parameter on the modification $\Delta\Gamma_{HHH}$
are systematically analyzed. Numerical results indicate that
$\Delta\Gamma_{HHH}$ deviates the one loop SM value substantially in the
selected parameter space formed by the mass of the top partner and
its mixing angle with the top partner.
\end{abstract}


\maketitle
\newpage

\section{introduction}
In the Standard Model (SM) of the particle physics,
the $SU_{L}(2)\times U_{Y}(1)$ symmetry is
broken by the vacuum expectation of the Higgs scalar field
introduced via the Higgs mechanism. Since the discovery of the Higgs
boson by the ATLAS~\cite{ATLAS:2012yve} and by the CMS Collaborations~\cite{CMS:2012qbp}
at the Large Hadron Collider (LHC) in 2012 ,
the observed properties of the Higgs boson are compatible with the
theoretical predictions, but up to now we know
little about the self couplings of the Higgs boson.
The self couplings of the Higgs boson, including
the trilinear and quartic self interaction which are respectively
denoted by $\lambda_{HHH}$ and $\lambda_{HHHH}$,
play a prominent role in determining the shape
of the Higgs potential in the SM. While the shape of the Higgs potential
is crucial not only in understanding the nature of
the electroweak symmetry breaking (EWSB) but also having profound implications
for cosmological stability~\cite{Markkanen:2018pdo, Horn:2020wif}.

The self coupling can be measured in the pair
production of the Higgs boson at the LHC
and the planned future colliders~\cite{CMS:2025hfp,
LinearColliderVision:2025hlt, Maura:2025rcv}.
So far there are no evident signals of the pair production of the Higgs
boson detected at the LHC, only  the trilinear self-coupling
modifier $\kappa_{\lambda}$ is reported and continuously
updated~\cite{ATLAS:2018dpp, ATLAS:2022jtk, ATLAS:2023qzf,
ATLAS:2024lsk, ATLAS:2024pov, ATLAS:2023gzn,
ATLAS:2023elc, ATLAS:2024lhu, ATLAS:2024ish,
ATLAS:2025hhd, CMS:2020tkr, CMS:2022kdx, CMS:2024rgy, CMS:2022hgz,
CMS:2022cpr, CMS:2024fkb, CMS:2024awa, CMS:2026znu, ATLAS:2026zrn}.
The modifier is defined as $\kappa_{\lambda}=\lambda^{eff}_{HHH}/\lambda_{HHH}$,
where $\lambda^{eff}_{HHH}$ and $\lambda_{HHH}$ are the effective
trilinear self coupling and the tree level SM value of this parameter, respectively.
The experimental advancements in recent years and the foreseeable future
motivate the evaluation of the $\lambda_{HHH}$ beyond the
tree level approximation. On the other hand, $\lambda_{HHH}$ can
serve as a well laboratory to test new physics beyond the SM since
the deviation of $\lambda_{HHH}$ from its SM value would point to
new physics. Therefore, a investigation
on the trilinear self coupling beyond the tree level
is necessary.

The one loop corrections to $\lambda_{HHH}$
have been widely studied in the SM and the minimal supersymmetric
standard model (MSSM) as well as the SM with an extended Higgs
sector~\cite{Belyaev:2002ua, Baur:2002rb, Baur:2003gp, Moretti:2004wa,
Brucherseifer:2013qva, Cao:2015oxx, Kanemura:2015fra, He:2016sqr, DiVita:2017eyz,
Braathen:2019pxr, Li:2019jba, Bahl:2022jnx,
Falaki:2023tyd, Stylianou:2023tgg, Abouabid:2024gms, Heinemeyer:2024hxa,
Braathen:2026glt}.
Recently, the tool \textsf{anyH3}, which is devoted to precise predictions for
the trilinear self coupling of the Higgs boson in any renormalizable model, is presented
in Ref.~\cite{Bahl:2023eau}. Thus to a large extent the one loop corrections to
$\lambda_{HHH}$ may be regarded as solved. However, the one loop corrections generated by
the models of vector-like quarks (VLQs) to $\lambda_{HHH}$ are still lacking.
Since its proposal decades ago~\cite{delAguila:1985mk,
Branco:1986my, Fishbane:1985gu}, the VLQs draw much attentions
as one of the new physics candidates~\cite{ Aguilar-Saavedra:2002phh, Dawson:2012di,
Dawson:2012mk, Aguilar-Saavedra:2013qpa, Cacciapaglia:2017gzh, Carvalho:2018jkq,
Arsenault:2022xty, Zhang:2023bsr, Alves:2023ufm}.
The most simplified VLQs model contains a colored Dirac fermion
which transforms as $T(3, 1, 2/3, q)$ under
$SU_{C}(3)\times SU_{L}(2)\times U_{Y}(1)\times U'(1)$. Quarks of this
type are often referred to as top partner which will contribute to $\lambda_{HHH}$
at one loop approximation. So we can exploit the effective trilinear self coupling
to test the vector-like quark model. Moreover, probing the VLQs
is one of the goals of the LHC experiments~\cite{ATLAS:2012isv, CMS:2017asf,
CMS:2018wpl, CMS:2019eqb, CMS:2022fck, ATLAS:2022hnn, ATLAS:2022tla,
ATLAS:2024gyc, ATLAS:2024zlo, Alves:2023ufm, ATLAS:2024fdw, CMS:2024bni}.
Consequently, it will get a double advantage to probe
the vector-like quark through observing the Higgs pair
production at the LHC. With this motivation, in this paper we will
systematically evaluate the one loop corrections to $\lambda_{HHH}$ in the
SM extended by the singlet vector-like quark model.
In principle, the top partner may mix with each generation of the SM quarks,
but the mixing with the first two generations is highly
restricted by the precision electroweak data and the flavor-changing
neutral currents processes~\cite{Aguilar-Saavedra:2002phh}.
For this reason, we assume that the top
partner only mixes with the SM top quark. Since at present there
is no experimental information on the quartic self
coupling, we do not consider it in this work.

However, there is complexity in evaluating the one loop corrections
to $\lambda_{HHH}$ due to the indefinite momentum of the Higgs lines.
A practical way out of this trouble is the effective potential method (also
known as the zero momentum approximation) by which the momentum of the
three Higgs boson lines are set to zero. This approach
has been applied in Ref.~\cite{Senaha:2018xek} and well results are obtained.
Recently, this method was utilized in evaluating two loop corrections to
$\lambda_{HHH}$ and $\lambda_{HHHH}$~\cite{Bahl:2025wzj}.
Thus in this paper we will adopt the zero momentum
approximation in our evaluation.

The paper is structured as follows. In section~\ref{SMoneloopevaluation},
the complete one loop corrections to $\lambda_{HHH}$ in the SM are evaluated.
In section~\ref{includingtoppartner}, the analytic one loop
contributions of the top parter to $\Gamma_{HHH}$ are displayed, effects of
the mass of the top partner and the mixing with the
top quark to $\Gamma_{HHH}$ are presented.
We draw our conclusion in section~\ref{summarysect}.

\section{one loop correction to the in the SM } \label{SMoneloopevaluation}

\begin{figure}
\begin{center}
\includegraphics[scale=0.50]{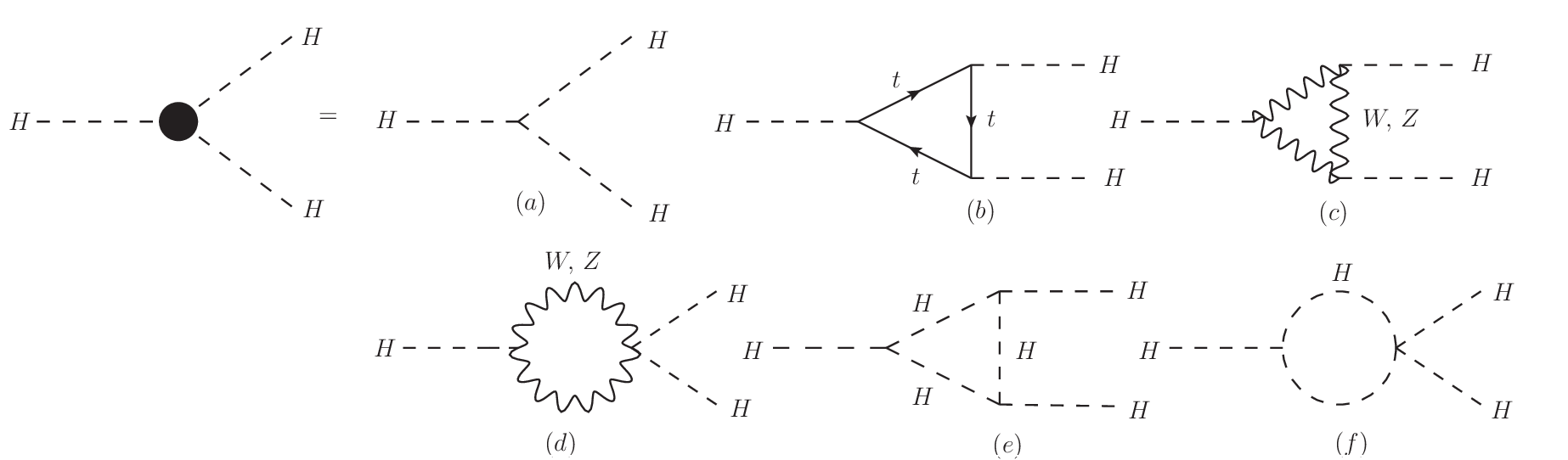}
\caption{The effective trilinear self coupling of
the Higgs boson up to one loop corrections in the SM:
(a) tree level coupling, (b) the top quark loop correction,
(c) and (d) $W$ or $Z$ boson loop corrections, (e) and (f) the
Higgs boson self coupling corrections. }   \label{SMoneloopdiagrams}
\end{center}
\end{figure}

In light of the zero momentum approximation,
the effective trilinear self coupling can be expanded as
\begin{equation}
\Gamma_{HHH}(p_{1}^{2},p_{2}^{2},p_{3}^{2})=
\Gamma_{HHH}(0,0,0)
+\mathcal{O}(p_{1}^{2},p_{2}^{2},p_{3}^{2})+\cdots,  \label{zeromomentumapprox}
\end{equation}
As shown in Refs.~\cite{Kanemura:2004mg, Kanemura:2002vm},
addition to a threshold enhancement occurring
when the incoming momentum is more than twice of the masses of the particle
propagating in the loop, the dominant contributions
to $\Gamma_{HHH}(p_{1}^{2},p_{2}^{2},p_{3}^{2})$ stem from
the momentum-independent part in Eq.~(\ref{zeromomentumapprox}).
Therefore, the evaluation will be greatly simplified.

Up to one loop approximation, the diagrams\footnote{All the
Feynman diagrams in this paper are produced
by \textsf{Jaxodraw}~\cite{Binosi:2003yf, Binosi:2008ig}.} that contribute
to the effective trilinear self coupling are
presented in Fig.~\ref{SMoneloopdiagrams}. Diagram (a) is the tree level
coupling, it reads
\begin{equation}
\Gamma_{a}=\Gamma^{{\rm{SM}, tree}}=\frac{3m^{2}_{{\rm{H}}}}{v}, \label{couplingoftreelevel}
\end{equation}
where $v=246\,{\rm{GeV}}$ is the vacuum expectation value.
We may classify the remaining diagrams into three types according to
the internal lines forming the triangle. The first one is diagram (b)
which is the top quark loop
correction, the second and the third types contain diagrams (c) and (d)
as well as diagrams (e) and (f), respectively.
Using dimensional regularization~\cite{tHooft:1972tcz, tHooft:1973mfk},
in the $D=4-2\varepsilon$ convention, the top quark loop correction in
can be conveniently evaluated
\begin{eqnarray}
\Gamma^{{\rm{SM}}}_{b}&=&-\lambda^{3}_{Htt}\int\frac{\D^{4}k}{(2\pi)^{4}}
\frac{{\rm{tr}}[(\slashed{k}+m_{t})(\slashed{k}+m_{t})
(\slashed{k}+m_{t})]}{(k^{2}-m^{2}_{t}+i\epsilon)^{3}}\nonumber\\
&=&-4m_{t}\lambda^{3}_{Htt}\times \frac{i}{(4\pi)^{2}}
\Big\{3\Big[\frac{1}{\varepsilon}
-\gamma_{E}+\ln(4\pi)-\ln\frac{m_{t}^{2}}{\mu^{2}}\Big]
-2+\mathcal{O}(\varepsilon)\Big\}, \label{SMbottomandtopcorrec}
\end{eqnarray}
where $\gamma_{E}=0.5772\dots$ is Euler-Mascheroni constant, and
$\mu$ is the renormalization scale.

The results from diagrams (c) and (d) are
\begin{eqnarray}
\Gamma^{{\rm{SM}}}_{c}&=&4\Big[\lambda^{3}_{HWW}\int\frac{\D^{4} k}{(2\pi)^{4}}
\frac{1}{(k^{2}-m_{W}^{2}+i\epsilon)^{3}}
+\lambda^{3}_{HZZ}\int\frac{\D^{4} k}{(2\pi)^{4}}
\frac{1}{(k^{2}-m_{Z}^{2}+i\epsilon)^{3}}\Big]\nonumber\\
&=&-\frac{2i}{(4\pi)^{2}}\big[\frac{\lambda^{3}_{HWW}}{m^{2}_{W}}
+\frac{\lambda^{3}_{HZZ}}{m^{2}_{Z}}+\mathcal{O}(\varepsilon)\big],\nonumber\\
\Gamma^{{\rm{SM}}}_{d}&=&4\Big[\lambda_{HWW}\lambda_{HHWW}\int\frac{\D^{4} k}{(2\pi)^{4}}
\frac{1}{(k^{2}-m_{W}^{2}+i\epsilon)^{2}}
+\lambda_{HZZ}\lambda_{HHZZ}\int\frac{\D^{4} k}{(2\pi)^{4}}
\frac{1}{(k^{2}-m_{Z}^{2}+i\epsilon)^{3}}\Big]\nonumber\\
&=&\frac{4i}{(4\pi)^{2}}
\Big\{\lambda_{HWW}\lambda_{HHWW}\Big[\frac{1}{\varepsilon}
-\gamma_{E}+\ln (4\pi)+\ln\frac{m^{2}_{W}}{\mu^{2}}\Big]\nonumber\\
&+&\lambda_{HZZ}\lambda_{HHZZ}\Big[\frac{1}{\varepsilon}-\gamma_{E}+\ln (4\pi)
+\ln\frac{m^{2}_{Z}}{\mu^{2}}\Big]
+\mathcal{O}(\varepsilon)\Big\}, \label{SMvectorbosoncorrec}
\end{eqnarray}

The corrections introduced by diagrams (e) and (f) are given by
\begin{eqnarray}
\Gamma^{{\rm{SM}}}_{e}&=&\lambda_{HHH}^{3}\int\frac{\D^{4} k}{(2\pi)^{4}}
\frac{1}{(k^{2}-m_{H}^{2}+i\epsilon)^{3}}\nonumber\\
&=&-\frac{i}{2(4\pi)^{2}}\frac{\lambda^{3}_{HHH}}{m_{H}^{2}}
+\mathcal{O}(\varepsilon), \label{SMtrilinearcorrc}
\end{eqnarray}
and
\begin{eqnarray}
\Gamma^{{\rm{SM}}}_{f}&=&\lambda_{HHH}\lambda_{HHHH}
\int\frac{\D^{4} k}{(2\pi)^{4}}
\frac{1}{(k^{2}-m_{H}^{2}+i\epsilon)^{2}}\nonumber\\
&=&\frac{i}{(4\pi)^{2}}\lambda_{HHH}\lambda_{HHHH}
\Big[\frac{1}{\varepsilon}-\gamma_{E}+\ln(4\pi)
-\ln\frac{m_{H}^{2}}{\mu^{2}}
+\mathcal{O}(\varepsilon)\Big],  \label{SMquarticcorrection}
\end{eqnarray}

By using the modified minimal subtraction ($\overline{{\rm{MS}}}$) scheme,
taking into account the crossing symmetry and summation over the color
of the top quark, we get the trilinear self coupling of the Higgs boson in the
SM
\begin{eqnarray}
\Gamma^{{\rm{SMoneloop}}}_{HHH}&=&\lambda_{HHH}+
\frac{1}{(4\pi)^{2}}\Big[-8N_{c}m_{t}
\lambda_{Htt}(2+3\ln\frac{m_{t}^{2}}{\mu^{2}})
+4\Big(\frac{\lambda^{3}_{HWW}}{m_{W}^{2}}
+\frac{\lambda^{3}_{HZZ}}{m_{Z}^{2}}\Big)\nonumber\\
&-&12\Big(\lambda_{HWW}\lambda_{HHWW}\ln\frac{m_{W}^{2}}{\mu^{2}}
+\lambda_{HZZ}\lambda_{HHZZ}\ln\frac{m_{Z}^{2}}{\mu^{2}}\Big)\nonumber\\
&+&\frac{3\lambda^{3}_{HHH}}{2m_{H}^{2}}
+3\lambda_{HHH}\lambda_{HHHH}
\ln\frac{m_{H}^{2}}{\mu^{2}}\Big].  \label{completeSMoneloopcorre}
\end{eqnarray}
All types of SM Higgs boson couplings
arise in Eq.~(\ref{completeSMoneloopcorre}), thus it is convenient to
investigate the effects of these couplings on the trilinear self coupling.
The couplings of the Higgs boson to top quark, massive gauge bosons as
well as the self couplings are~\cite{Djouadi:2005gi}
\begin{eqnarray}
\lambda_{Htt}&=&\frac{m_{t}}{v}=(\sqrt{2}G_{f})^{1/2}m_{t},\nonumber\\
\lambda_{HVV}&=&\frac{2m^{2}_{V}}{v}=2(\sqrt{2}G_{f})^{1/2}m_{V}^{2},\quad
\lambda_{HHVV}=\frac{2m^{2}_{V}}{v^{2}}
=2\sqrt{2}G_{f}m_{V}^{2},\quad V=W,\,Z \nonumber\\
\lambda_{HHH}&=&\frac{3m^{2}_{H}}{v}=3(\sqrt{2}G_{f})^{1/2}m_{H}^{2},\quad
\lambda_{HHHH}=\frac{3m^{2}_{H}}{v^{2}}=3\sqrt{2}G_{f}m_{H}^{2}.  \label{coplingsofthehiggs}
\end{eqnarray}
where $G_{f}=1.664\times10^{-5}\,{\rm{GeV}}^{-2}$ is the Fermi coupling
constant. The masses of the top quark, gauge bosons and the Higgs boson
are~\cite{ParticleDataGroup:2026}
\begin{eqnarray}
m_{t}&=&172.60\pm 0.27\,{\rm{GeV}},\quad
m_{W}=80.3625\pm 0.0077\,{\rm{GeV}},\nonumber\\
m_{Z}&=&91.1879\pm 0.0020\,{\rm{GeV}},\quad
m_{H}=125.13\pm 0.11\,{\rm{GeV}}. \label{massesoftwzandthehiggs}
\end{eqnarray}
We define the following ratio in the SM
\begin{equation}
\Delta\Gamma_{HHH}^{{\rm{SM}}}=\frac{\Gamma^{{\rm{SMoneloop}}}_{HHH}-\Gamma^{{\rm{SM}, tree}}_{HHH}}
{\Gamma^{{\rm{SM}, tree}}_{HHH}}, \label{ratioinSM}
\end{equation}
By employing the parameters in Eq.~(\ref{massesoftwzandthehiggs}),
we find the ratio of the
trilinear self coupling in the SM up to one loop correction
in the zero momentum approximation at the scale $\mu=m_{t}$
\begin{equation}
\Delta\Gamma_{HHH}^{{\rm{SM}}}= -0.08 \label{resultsinthesm}
\end{equation}
This result indicates that
the one loop correction plays the role in decreasing $\Gamma_{HHH}$
by the amount about $10\%$ compared with the tree level value.
Since we just retain the first term
in Eq.~(\ref{zeromomentumapprox}) and neglect the contribution of other
quarks in the SM, the results in Eq.~(\ref{resultsinthesm}) should be
perceived as a lowest reduction up to one loop evaluation.
If we incorporate higher order terms in Eq.~(\ref{zeromomentumapprox})
or extend it to a two loop evaluation, it is expected the ratio
$\Delta\Gamma_{HHH}^{{\rm{SM}}}$ will be reduced more than the value
in Eq.(\ref{resultsinthesm}). As we know the singlet top partner
can couple to the Higgs boson so that new contributions will be
generated to the effective trilinear self coupling
by the coupling to the top partner. The LHC experiments
indicate that the lower limit of the mass for the top partner is greater
than $1200\,{\rm{GeV}}$~\cite{ATLAS:2024zlo, Cingiloglu:2023ylm, ATLAS:2025bzt},
and the coupling of the Higgs boson to the top partner is proportional
to the mass of the top partner, thus it is expected that the
results in Eq.~(\ref{resultsinthesm}) will be changed
if the corrections of the top partner are included in the evaluation.
This will be presented in the subsequent sections.

\section{including the contribution of singlet vector-like top partner} \label{includingtoppartner}

\begin{figure}
\begin{center}
\includegraphics[scale=0.70]{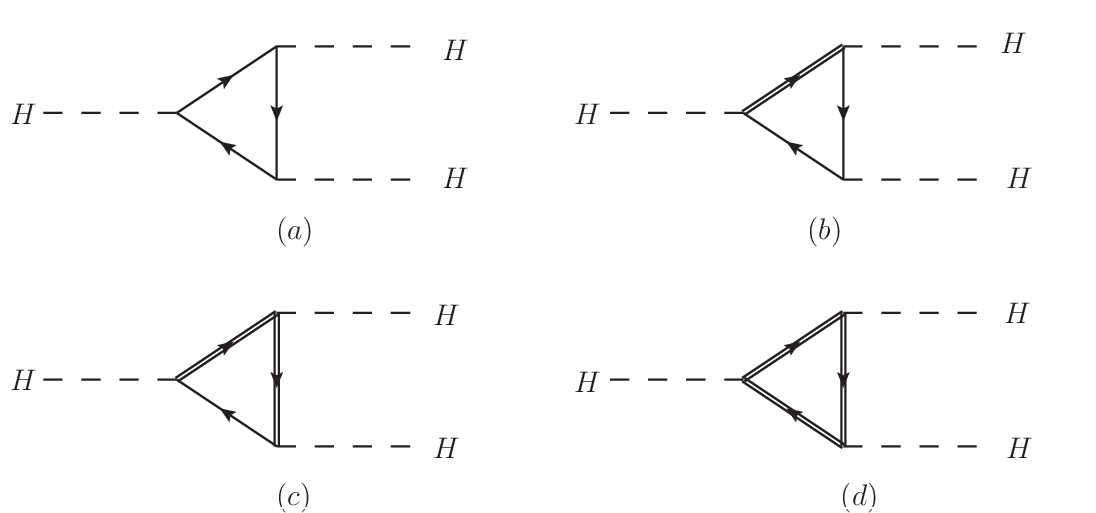}
\caption{The one loop contrbutions of the top partner to the triple Higgs
self-interaction. Double lines denote the vector-like top partner.
Diagram (a), new top quark loop correction to the trilinear coupling,
(b), (c) and (d) denote one loop correction to $\lambda_{HHH}$ with one, two and three
top partner internal lines, respectively. }         \label{oneloopincludingtoppartner}
\end{center}
\end{figure}

In this section, we consider the contributions of the singlet vector-like
top partner to the effective trilinear self coupling.
The mixing between the top quark and its singlet partner is~\cite{Dawson:2012di}
\begin{equation}
\begin{pmatrix}
t_{L}\\
T_{L}
\end{pmatrix}
=\begin{pmatrix}
\cos\theta_{L}&-\sin\theta_{L}\\
\sin\theta_{L}&\cos\theta_{L}
\end{pmatrix}
\begin{pmatrix}
\mathcal{T}^{1}_{L}\\
\mathcal T^{2}_{L}
\end{pmatrix},\quad\quad
\begin{pmatrix}
t_{R}\\
T_{R}
\end{pmatrix}
=\begin{pmatrix}
\cos\theta_{R}&-\sin\theta_{R}\\
\sin\theta_{R}&\cos\theta_{R}
\end{pmatrix}
\begin{pmatrix}
\mathcal{T}^{1}_{R}\\
\mathcal T^{2}_{R}
\end{pmatrix},  \label{topandthepartnermixing}
\end{equation}
where $t_{L, R}$ and $T_{L,R}$ are the respective
mass eigenstates of top quark and its vector-like singlet partner,
while $\mathcal{T}^{1}_{R, L}$ and $\mathcal{T}^{2}_{R, L}$
are the corresponding weak eigenstates.
The couplings to the SM Higgs boson are~\cite{Dawson:2012di}
\begin{equation}
\mathcal{L}=-\frac{m_{t}}{v}c_{tt}\overline{t_{L}}t_{R}h
-\frac{M_{T}}{v}c_{TT}\overline{T_{L}}T_{R}h
-\frac{M_{T}}{v}c_{tT}\overline{t_{L}}T_{R}h
-\frac{m_{t}}{v}c_{Tt}\overline{T_{L}}t_{R}h+\rm{h.c.}, \label{lagrangianofhiggsandTt}
\end{equation}
with the coefficients given by
\begin{equation}
c_{tt}=c_{L}^{2},\quad
c_{TT}=s_{L}^{2},\quad c_{tT}=c_{Tt}=s_{L}c_{L},  \label{definitionofclsl}
\end{equation}
where $s_{L}$ and $c_{L}$ are the abbreviation
of $\sin\theta_{L}$ and $\cos\theta_{L}$, respectively.

By making use the vertexes listed in Ref.~\cite{Zhang:2023bsr} which are derived
from the Lagrangian in Eq.~(\ref{lagrangianofhiggsandTt}),
we list new contributions induced by the four
diagrams in Fig.~\ref{oneloopincludingtoppartner}
\begin{eqnarray}
\Gamma^{{\rm{VLQ}}}_{a}&=&\Big(\frac{m_{t}c_{tt}}{v}\Big)^{3}
\int\frac{\D^{4}k}{(2\pi)^{4}}
\frac{{\rm{tr}}[(\slashed{k}+m_{t})(\slashed{k}+m_{t})
(\slashed{k}+m_{t})]}{(k^{2}-m_{t}^{2}+i\epsilon)^{3}}\nonumber\\
&=&\frac{i}{(4\pi)^{2}}4m_{t}\Big(\frac{m_{t}c_{tt}}{v}\Big)^{3}
\Big[3\Big(\frac{1}{\varepsilon}-\gamma_{E}
+\ln(4\pi)-\ln\frac{m^{2}_{t}}{\mu^{2}}\Big)
-2+\mathcal{O}(\varepsilon)\Big], \label{onetoppartnerloop}
\end{eqnarray}
\begin{eqnarray}
\Gamma^{{\rm{VLQ}}}_{b}&=&\frac{M_{T}m_{t}^{2}c_{tt}c_{tT}^{2}}{v^{3}}
\int\frac{\D^{4}k}{(2\pi)^{4}}
\frac{{\rm{tr}}[(\slashed{k}+m_{t})(\slashed{k}+M_{T})(\slashed{k}+m_{t})]}
{(k^{2}-M_{T}^{2}+i\epsilon)(k^{2}-m_{t}^{2}+i\epsilon)^{2}}\nonumber\\
&=&\frac{i}{(4\pi)^{2}}\frac{4M_{T}m_{t}^{2}c_{tt}c_{tT}^{2}}{v^{3}}
\Big\{(2m_{t}+M_{T})\Big[\frac{1}{\varepsilon}
-\gamma_{E}+\ln(4\pi)+\Big(1-\ln\frac{m_{t}^{2}}{\mu^{2}}
-\frac{m_{T}^{2}}{m_{t}^{2}-M_{T}^{2}}
\ln\frac{m_{t}^{2}}{M_{T}^{2}}\Big)\Big]\nonumber\\
&-&\frac{2m_{t}^{2}(m_{t}+M_{T})}{m_{t}^{2}-M_{T}^{2}}
\Big(1-\frac{M_{T}^{2}}{m_{t}^{2}-m_{T}^{2}}
\ln\frac{m_{t}^{2}}{M_{T}^{2}}\Big)+\mathcal{O}(\varepsilon)\Big\}, \label{twotoppartnerloop}
\end{eqnarray}
\begin{eqnarray}
\Gamma^{{\rm{VLQ}}}_{c}&=&\frac{m_{t}M^{2}_{T}c_{TT}c_{tT}^{2}}{v^{3}}
\int\frac{\D^{4}k}{(2\pi)^{4}}
\frac{{\rm{tr}}[(\slashed{k}+M_{T})(\slashed{k}+M_{T})(\slashed{k}+m_{t})]}
{(k^{2}-m_{t}^{2}+i\epsilon)(k^{2}-M_{T}^{2}+i\epsilon)^{2}}\nonumber\\
&=&\frac{i}{(4\pi)^{2}}\frac{4m_{t}M^{2}_{T}c_{TT}c_{tT}^{2}}{v^{3}}
\Big\{(m_{t}+2M_{T})\Big[\frac{1}{\varepsilon}
-\gamma_{E}+\ln(4\pi)+\Big(1-\ln\frac{M_{T}^{2}}{\mu^{2}}
-\frac{m_{t}^{2}}{M_{T}^{2}-m_{t}^{2}}
\ln\frac{M_{T}^{2}}{m_{t}^{2}}\Big)\Big]\nonumber\\
&-&\frac{2M_{T}^{2}(m_{t}+M_{T})}{M_{T}^{2}-m_{t}^{2}}
\Big(1-\frac{m_{t}^{2}}{M_{T}^{2}-m_{t}^{2}}
\ln\frac{M_{T}^{2}}{m_{t}^{2}}\Big)+\mathcal{O}(\varepsilon)\Big\}, \label{threetoppartnerloop}
\end{eqnarray}
\begin{eqnarray}
\Gamma^{{\rm{VLQ}}}_{d}&=&\Big(\frac{M_{T}c_{TT}}{v}\Big)^{3}
\int\frac{\D^{4}k}{(2\pi)^{4}}
\frac{{\rm{tr}}[(\slashed{k}+M_{T})(\slashed{k}+M_{T})
(\slashed{k}+M_{T})]}{(k^{2}-M_{T}^{2}+i\epsilon)^{3}}\nonumber\\
&=&\frac{i}{(4\pi)^{2}}4M_{T}\Big(\frac{M_{T}c_{TT}}{v}\Big)^{3}
\Big[3\Big(\frac{1}{\varepsilon}-\gamma_{E}
+\ln(4\pi)-\ln\frac{M^{2}_{T}}{\mu^{2}}\Big)
-2+\mathcal{O}(\varepsilon)\Big], \label{newtopquarkloop}
\end{eqnarray}

Taking into account the crossing symmetry and summation over the color,
we get the effective trilinear self coupling with inclusion of contribution of
the top partner in the $\overline{{\rm{MS}}}$ scheme
\begin{eqnarray}
\Gamma^{{\rm{SM+VLQ}}}_{HHH}&=&\lambda_{HHH}+
\frac{1}{(4\pi)^{2}}\Big\{-8N_{c}m_{t}
\lambda_{Htt}(2+3\ln\frac{m_{t}^{2}}{\mu^{2}})
+4\Big(\frac{\lambda^{3}_{HWW}}{m_{W}^{2}}
+\frac{\lambda^{3}_{HZZ}}{m_{Z}^{2}}\Big)\nonumber\\
&-&12\Big(\lambda_{HWW}\lambda_{HHWW}\ln\frac{m_{W}^{2}}{\mu^{2}}
+\lambda_{HZZ}\lambda_{HHZZ}\ln\frac{m_{Z}^{2}}{\mu^{2}}\Big)
+\frac{3\lambda^{3}_{HHH}}{2m_{H}^{2}}\nonumber\\
&+&3\lambda_{HHH}\lambda_{HHHH}
\ln\frac{m_{H}^{2}}{\mu^{2}}+8N_{c}m_{t}\Big(\frac{m_{t}c_{tt}}{v}\Big)^{2}
\Big(2+3\ln\frac{m_{t}^{2}}{\mu^{2}}\Big)\nonumber\\
&-&\frac{8N_{c}M_{T}m_{t}^{2}c_{tt}c_{tT}^{2}}{v^{3}}
\Big[(2m_{t}+M_{T})\Big(1-\ln\frac{m_{t}^{2}}{\mu^{2}}
-\frac{M_{T}^{2}}{m_{t}^{2}-M_{T}^{2}}
\ln\frac{m_{t}^{2}}{M_{T}^{2}}\Big)\nonumber\\
&-&\frac{2m_{t}^{2}(m_{t}+M_{T})}{m_{t}^{2}-M_{T}^{2}}
\Big(1-\frac{M_{T}^{2}}{m_{t}^{2}-M_{T}^{2}}
\ln\frac{m_{t}^{2}}{M_{T}^{2}}\Big)\Big]\nonumber\\
&-&\frac{8N_{c}m_{t}M^{2}_{T}c_{TT}c_{tT}^{2}}{v^{3}}
\Big[(m_{t}+2M_{T})\Big(1-\ln\frac{M_{T}^{2}}{\mu^{2}}
-\frac{m_{t}^{2}}{M_{T}^{2}-m_{t}^{2}}
\ln\frac{M_{T}^{2}}{m_{t}^{2}}\Big)\nonumber\\
&-&\frac{2M_{T}^{2}(m_{t}+M_{T})}{M_{T}^{2}-m_{t}^{2}}
\Big(1-\frac{m_{t}^{2}}{M_{T}^{2}-m_{t}^{2}}
\ln\frac{M_{T}^{2}}{m_{t}^{2}}\Big)\Big]\nonumber\\
&+&4N_{c}M_{T}\Big(\frac{M_{T}c_{TT}}{v}\Big)^{3}
\Big(2+3\ln\frac{M^{2}_{T}}{\mu^{2}}\Big)\Big\}. \label{totaltoppartnercontr}
\end{eqnarray}
We define the ratio
\begin{equation}
\Delta\Gamma_{HHH}^{{\rm{SM+VLQ}}}=\frac{\Gamma^{{\rm{SM+VLQ}}}_{HHH}
-\Gamma^{{\rm{SMoneloop}}}_{HHH}}
{\Gamma^{{\rm{SMoneloop}}}_{HHH}}, \label{ratioinSMandVLQ}
\end{equation}
to measure the effects of singlet vector-like top partner on the
the trilinear self coupling. The numerical results will be presented in
the next section.

\section{results and discussions} \label{resultsanddiscussion}

\subsection{choice of the parameters}   \label{parametersinnumericaleval}
\begin{figure}
\begin{center}
\includegraphics[scale=1.50]{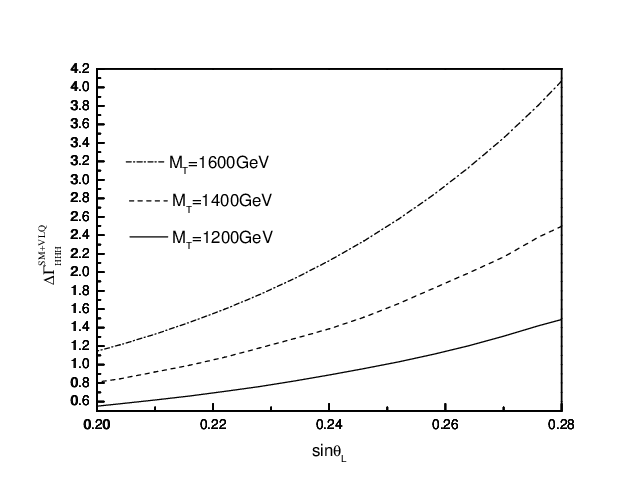}
\caption{The ratio $\Delta \Gamma^{{\rm{SM+VLQ}}}_{HHH}$ as a function of the mixing parameters for
fixed mass of the top partner.}  \label{fixedMTandmiixng}
\end{center}
\end{figure}

\begin{figure}
\begin{center}
\includegraphics[scale=1.50]{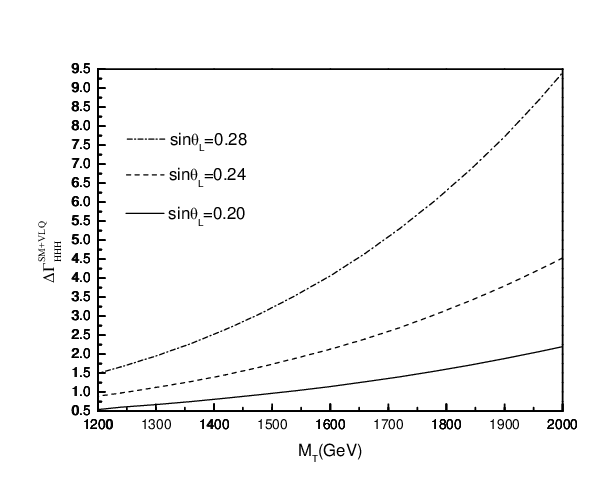}
\caption{The ratio $\Delta \Gamma^{{\rm{SM+VLQ}}}_{HHH}$ as a function of mass of the
top partner for fixed mixing parameters.}         \label{fixedmixingangles}
\end{center}
\end{figure}

Before performing the numerical evaluation, two inputting parameters
introduced by the singlet top partner model must be clarified.
The first one is the mass of the top partner. Although the ATLAS and the CMS
collaborations searched for VLQs\footnote{Addition to the singlet vector-like model
investigated in this paper, there are doublet and triplet vector-like
quark models~\cite{Aguilar-Saavedra:2013qpa, Alves:2023ufm}.} for several
years through the decays of the VLQs to the SM particles,
up to now there is no definite signals observed,
only the lower limit of the masses of the VLQs
are established~\cite{ ATLAS:2024zlo, Cingiloglu:2023ylm, ATLAS:2025bzt}.
Based on the direct searches by the ATLAS and CMS, assuming the VLQs decay only to
the third generation quarks, the mass limit $M_{T}>1.27$\,TeV for the singlet vector-like
top quark is obtained in Ref.~\cite{Cingiloglu:2023ylm}. Using the
data recorded by the ATLAS detector from 2015 to 2018 during the Run 2 of the LHC,
the VLQ mass below 1150 GeV is excluded for the SU(2) singlet model~\cite{ATLAS:2024zlo}.
As stated in Ref.~\cite{ATLAS:2025bzt}, the strongest observed limit for the mass of the
singlet vector-like top quark is $M_{T}>1280$\,GeV which is set by the search
for vector-like T and B quark pairs in final states
with leptons at $\sqrt{s}=13$\,TeV by the CMS collaboration~\cite{CMS:2018zkf}.
To balance these results, we take $M_{T}=1200$\,GeV as the lower limit
in our numerical evaluation.

Another one is the mixing parameter $\sin\theta_{L}$.
This quantity can be determined indirectly by the
loop-induced diphoton decay or gluon fusion processes
of the Higgs boson to which the singlet top partner
may contribute~\cite{Dawson:2012di, Dawson:2013uqa, Arhrib:2016rlj}.
According to the exploration of the effects
of the singlet top partner on the Higgs
couplings~\cite{Aguilar-Saavedra:2006uim, Xiao:2014kba},
an upper limit $\sin\theta_{L}<0.40$ is derived from the
combined $H\rightarrow gg$ and $H\rightarrow \gamma\gamma$
cross section and branching ratio, respectively.
In fact, the determination of $\sin\theta_{L}$ is
closely related to the mass of the top partner
in specific processes. For instance, using the
different branching ratios of $H\rightarrow b\bar{s}$ as
the constraint and assuming the mass of the singlet top partner
is not more than 2000\,GeV, an upper limit $0.24$ is
obtained~\cite{Zhang:2023bsr}. In our numerical evaluation,
we set $\sin \theta_{L}=0.28$ as the upper limit of the mixing
parameter in our analysis.

\subsection{numerical results and discussions}

In order to investigate the effects of the top partner
on the $\Delta \Gamma^{{\rm{SM+VLQ}}}_{HHH}$, we set the mass of the top parter to be
$1200\,{\rm{GeV}}$, $1400\,{\rm{GeV}}$ and $1600\,{\rm{GeV}}$, respectively, The variations
of $\Delta \Gamma^{{\rm{SM+VLQ}}}_{HHH}$ with the mixing parameter $\sin\theta_{L}$
are displayed in Fig.~\ref{fixedMTandmiixng}.
It is evident that the ratio $\Delta \Gamma^{{\rm{SM+VLQ}}}_{HHH}$ is sensitive to the value of
$\sin\theta_{L}$, there is obvious variation of $\Delta \Gamma^{{\rm{SM+VLQ}}}_{HHH}$
along with the increasing of $\sin\theta_{L}$. The values of
$\Delta \Gamma^{{\rm{SM+VLQ}}}_{HHH}$ are positive definite
which is contrary to the case of SM. This indicates that in the
parameter space selected the contributions of top partner
can significantly enhance the effective trilinear coupling of the Higgs boson.
The greater value the top partner it takes,
$\Delta \Gamma^{{\rm{SM+VLQ}}}_{HHH}$ will grow more rapidly.
For instance, if $M_{T}=1600$\,GeV, the lower bound of
$\Delta \Gamma^{{\rm{SM+VLQ}}}_{HHH}$ is greater than
the upper limit when $M_{T}=1200$\,GeV.

On the contrary, we may fix the mixing parameter to be $\sin\theta_{L}=0.20$,
$0.24$ and $0.28$, the variation of $\Delta \Gamma^{{\rm{SM+VLQ}}}_{HHH}$ along with the masses of the
top partner falling in the range $[1200,\, 2000]$\,GeV is presented
in Fig.~\ref{fixedmixingangles}. We can read from Fig.~\ref{fixedmixingangles} is that
$\Delta \Gamma^{{\rm{SM+VLQ}}}_{HHH}$ exhibits similar
trend as the case of fixed $M_{T}$ discussed in the above
paragraph. When the mixing parameter takes the lowest value $\sin\theta_{L}=0.20$,
$\Delta \Gamma^{{\rm{SM+VLQ}}}_{HHH}$ grows slowly in the whole mass range.
If $\sin\theta_{L}$ raises to $0.28$, $\Delta \Gamma^{{\rm{SM+VLQ}}}_{HHH}$
increases quickly when the mass of the vector like top partner beyond
$1600$\,GeV. However, in the experiments it is impossible to
observe the effective trilinear coupling
$\Gamma_{HHH}$ exclusively by single channel.
Thus as a rough estimate, according to the definition of $\kappa_{\lambda}$
and its newly value reported by ATLAS~\cite{ATLAS:2026zrn},
we conclude that a larger mixing parameter can
only match to a lower mass of the top partner
so that the calculated results based on SM and singlet vector-like
top partner can be consistent with the experiments.

\section{summary}  \label{summarysect}

In this work, by using the zero momentum approximation,
we firstly evaluate the full one loop corrections to the trilinear
self coupling of the Higgs boson in the SM.
At the scale of top quark mass, numerical result of
the ratio $\Delta\Gamma_{HHH}^{{\rm{SM}}}$ indicates
that the one loop correction play the role to lower
the effective trilinear self coupling of the Higgs boson
compared the tree level value. Then we extend the SM
by inclusion of a singlet vector-like top partner,
assuming the top partner only mixes with the top
quark, the one loop contribution of the top partner
to $\Gamma_{HHH}$ is obtained. The effects of the
mass of the top partner and the mixing parameter to the ratio
$\Delta \Gamma^{{\rm{SM+VLQ}}}_{HHH}$ are analyzed.
We find that the contribution of the top partner
significantly enhance the effective trilinear self coupling
of the Higgs boson, the ratio $\Delta \Gamma^{{\rm{SM+VLQ}}}_{HHH}$
is always positive in the parameter space selected. Further more,
the numerical results suggest that in order to get
reasonable results for $\Delta \Gamma^{{\rm{SM+VLQ}}}_{HHH}$,
the mass of the top partner and the mixing parameter can not be
taken too large value.

\section{acknowledgements}
The author thanks Hong-Ying Jin and
Tom G. Steele for valuable discussions.


\end{document}